\documentclass[twocolumn,showpacs,preprintnumbers,amsmath,amssymb]{revtex4}

\usepackage{graphicx}
\usepackage{dcolumn}
\usepackage{bm}
\usepackage{colortbl}

\begin{document}

\preprint{APS/*****}

\title{Mott-insulator to metal transition in filling-controlled SmMnAsO$_{1-x}$}

\author{Y.Shiomi$^{\, 1,2}$}
\author{S.Ishiwata$^{\, 1,3}$}
\author{Y.Taguchi$^{\, 2,3}$}
\author{Y.Tokura$^{\, 1,2,3,4}$}
\affiliation{$^{1}$
Department of Applied Physics, University of Tokyo, Tokyo 113-8656, Japan }
\affiliation{$^{2}$
Cross-Correlated Materials Research Group (CMRG), RIKEN Advanced Science Institute, Wako 351-0198, Japan
}
\affiliation{$^{3}$
Correlated Electron Research Group (CERG), RIKEN Advanced Science Institute, Wako 351-0198, Japan
}
\affiliation{$^{4}$
Multiferroics Project, ERATO, Japan Science and Technology Agency (JST), Tokyo 113-8656, Japan
}

\date{\today}

\begin{abstract}
Transport and magnetic properties have been systematically investigated for SmMnAsO$_{1-x}$ with controlled electron-doping. As the electron band-filling is increased with the increase of oxygen deficiency ($x$), the resistivity monotonically decreases and the transition from Mott-insulator to metal occurs between $x=0.17$ and $0.2$. Seebeck coefficient at room temperature abruptly changes around the critical doping level from a large value ($\sim -300$ ${\rm \mu}$V/K) to a small one ($\sim -50$ ${\rm \mu}$V/K) both with negative sign. In the metallic compounds with $x=0.2$ and $x=0.3$, Mn spins order antiferromagnetically around $30$ K, and the Hall coefficient with the negative sign shows a reduction in magnitude upon the magnetic transition, indicating the change in the multiple Fermi surfaces. Gigantic positive magnetoresistance effect is observed in a wide range of temperature, reaching up to $60$ \% at $2$ K for the $x=0.3$ compound. The effect is attributed to the field-induced change of the complex Fermi surfaces in this multi-orbital correlated electron system.          
\end{abstract}

\maketitle

\section{\label{sec:level1}Introduction}

Since the recent discovery of superconductivity in doped LaFeAsO \cite{kamihara,takahashi} and related materials with high transition critical temperatures, the Fe-based layered pnictides have attracted much interest similarly to the cuprate superconductors. The crystal structure of the parent material $R$FeAsO ($R$: rare earth metal) is classified into ZrCuSiAs-type (space group $P4/nmm$). Up to now, more than $70$ materials with $RMPn$O formula ($M$: transition metal, $Pn$: pnictogen) have been known to possess this crystal structure. The presence of such a large number of materials is due to the fact that the substitution of many other $3d$ or $4d$ transition metals are possible for $M$-site in addition to rich variation in $Pn$-site (P, As, Sb, and Bi) and $R$-site (La-Dy).\cite{Ozawa} In this crystal structure, insulating ($R^{3+}$O$^{2-}$)$^{+}$-layer and conducting ($M^{2+}\,Pn^{3-}$)$^{-}$-layer are alternatingly stacked along the $c$-axis. As an example, the structure of SmMnAsO is depicted in Fig. 1(a). In the (Sm$^{3+}$O$^{2-}$)$^{+}$-layer, O$^{2-}$ is tetrahedrally coordinated by Sm$^{3+}$ ions. The (Mn$^{2+}$As$^{3-}$)$^{-}$-layer also has the same connectivity of the ions, but the cation and anion are interchanged. \par

The physical properties of the compounds La$MPn$O ($M$=Mn, Co, and Ni, $Pn$=P and As) have been extensively studied by Hosono and coworkers. \cite{Hosono} According to their study, LaMnPO is an antiferromagnetic insulator with an optical gap larger than $1$ eV and high N\'eel temperature (above $375$K). LaCoPO and LaCoAsO are ferromagnetic metals with $T_{c}=43$ K and $70$K, respectively. LaNiPO and LaNiAsO are paramagnetic metals and show superconductivity in the low temperature region ($<3$ K). For LaMnPO, Yanagi $et\ al.$ synthesized a bulk polycrystalline sample and investigated the physical properties. \cite{Yanagi} The resistivity is $\sim 1$k${\rm \Omega}$cm around room temperature and shows insulating temperature dependence, which well obeys the Arrhenius-type relation. \cite{Yanagi} A neutron powder diffraction measurement indicates that the moment of Mn$^{2+}$ ($3d^5$) spin is $2.26\mu_{B}$, which is smaller than $5 \mu_{B}$/Mn expected for the high-spin state ($S=5/2$), and that the Mn spins within the MnP-layer show anti-parallel arrangement and parallel arrangement along the $c$-axis at room temperature. Quite recently, $R$MnPO with other rare earth element ($R$=Nd, Sm, and Gd) were also investigated. \cite{Yanagi-2} The resistivity at room temperature is about three orders of magnitude smaller than that of LaMnPO, but the temperature dependence remains semiconducting. Magnetization measurement indicates that Mn$^{2+}$ spins order antiferromagnetically, and that the $T_{N}$ is higher than $350$ K, similarly to LaMnPO. Concerning LaMnAsO, Kayanuma $et\ al.$  \cite{kayanuma} synthesized an epitaxially grown film with a small amount of impurity (LaMnO$_{3}$). The resistivity is $\sim 1 {\rm \Omega}$cm at room temperature, although a relatively large charge gap of $\sim 1$ eV is predicted by the band calculation. \cite{kayanuma} This material is also considered as an antiferromagnetic insulator analogous to LaMnPO, \cite{Hosono} yet the magnetic property has not been reported. 

\par

For $R$FeAsO, $R$-dependent physical peroperties have been intensively investigated, in which the superconducting transition temperature of doped materials depends significantly on the $R$ species. \cite{eisaki} In the case of $M=$Co, Ohta and Yoshimura \cite{yoshimura} recently investigated the magnetism of the $R$CoAsO ($R$=La-Gd), and showed that the antiferromagnetic state is more stable than the ferromagnetic state in the compounds with relatively small $R$-ions. On the other hand, there have been few reports on $R$MnAsO with other rare earth element than $R$=La and their physical properties are hardly unveiled. Recently, Marcinkova $et\ al.$ reported the magnetic property of NdMnAsO, \cite{Marcinkova} which is an antiferromagentic semiconductor with $T_{N}= 359$K. Neutron powder diffraction measurement showed that, below $T_{N}$, Mn$^{2+}$-spins order antiferromagnetically with pointing along $c$-axis with the same spin-ordering pattern as in LaMnPO. With further cooling, the Mn$^{2+}$-spins show reorientation, lying within the $ab$-plane at $T_{\rm SR}=23$ K, accompanied by the antiferrromagnetic order of Nd$^{3+}$ moments. Across the spin-reorientation transition, the ordering pattern of Mn$^{2+}$ spins remains unchanged. Nd$^{3+}$ moments align ferromagnetically within a NdO-layer and antiferromagnetically along the perpendicular direction. \par

A promising way to find the emergent phenomena in such correlated-electron materials is the band-filling (doping) control, which has been extensively adopted in the past decades, \cite{imada} and is also the main focus of the present paper. In $R$MnPO, the hole-doping by substitution of $R$ (La, Nd, Sm, and Gd) with Ca or Cu below $10$ \% was reported. \cite{Yanagi, Yanagi-2} While undoped $R$MnPO shows $n$-type conduction, the Ca- or Cu-doped sample has $p$-type carrier as judged from the Seebeck coefficient. The resistivity for the doped-samples decreases in magnitude, but the temperature dependence remains semiconducting. In this paper, we demonstrate that doping a large amount of electrons is possible for a Mn-based oxy-arsenide SmMnAsO$_{1-x}$ by introducing oxygen deficiency ($x$) with the use of high-pressure synthesis technique. Undoped SmMnAsO is a Mott-insulator, and the insulator-metal transition occurs at $x=0.2$. The metallic samples ($x=0.2$ and $0.3$) show antiferromagnetic transition at around $30$K, and the Hall coefficient shows a reduction upon the transition. These metallic state exhibits unconventionally large positive magnetoresistance up to $60$\% at $2$ K under $9$ T. The Seebeck coefficient with negative sign changes from a large absolute value arising from variable-range-hopping (VRH) conduction to a metallic small one upon the insulator-metal transition as a function of the oxygen deficiency ($x \sim 0.2$), ensuring the bulk nature of the transition. 

\begin{figure}[thbp]
\begin{center}
\includegraphics*[width=8cm]{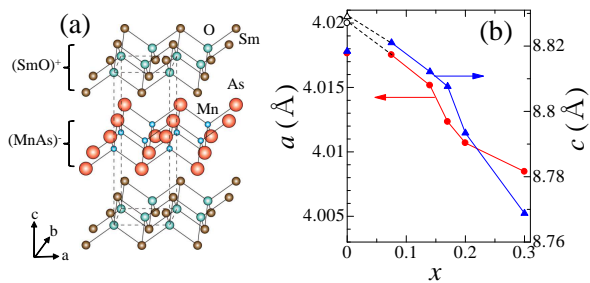}
\end{center}
\caption{(color online). (a) A schematic diagram of the crystal structure of SmMnAsO. Dashed lines show the tetragonal unit cell. (b) Lattice constants ($a$ and $c$) are plotted against nominal oxygen-deficiency ($x$).}
\label{fig1}
\end{figure}

\section{Experiment}

First, SmAs was prepared as a precursor by sintering Sm (3N, shot) and As (5N, powder) at 1050 $^\circ \mathrm{C}$ for $24$ h in an evacuated silica tube. SmMnAsO samples with controlled oxygen deficiency (SmMnAsO$_{1-x}$) were made by high-pressure synthesis technique with the use of a cubic anvil press. SmAs, Mn (3N, powder) and MnO (3N, powder) were mixed in a prescribed ratio ($\sim 0.2$ g) in an Ar-filled glove box and fired at 1150 $^\circ \mathrm{C}$ for 1 hour under 6.5GPa in a BN crucible. The products were examined with powder X-ray diffraction. The resistivity, Hall resistivity, and specific heat were measured by using a Physical Property Measurement System (Quantum Design). The magnetoresistance was measured in the longitudinal configuration, {\it i.e.} the magnetic field parallel to the electric current. Seebeck coefficient and thermal conductivity were measured with a conventional steady state method. Magnetization measurement was performed by using a SQUID magnetometer, Magnetic Property Measurement System (Quantum Design).

\section{Results and discussions}

\subsection{Sample characterization}

High-pressure synthesis technique enabled us to obtain the oxygen-deficient samples up to $x$=0.3. Almost single-phase samples were obtained up to $x$=0.17, but for $x=0.2$ and $0.3$, the products contained a small amount (less than 4\% in volume-fraction) of SmAs as an impurity. Also, the presence of a tiny amount of ferromagnetic impurity MnAs was indicated by the magnetization measurement although not identified by the powder X-ray diffraction. Figure \ref{fig1}(b) shows the obtained lattice constants ($a$ and $c$) for these samples. The open circle and triangle indicate the reported values for the $x$=0 compound synthesized by a different method. \cite{Ozawa} In the present sample preparation condition, the $c$ value of $x$=0 is smaller than that of $x=0.075$. It appears that the obtained samples contain more oxygen than the nominal value and the stability of Sm$_{2}$O$_{3}$ phase prevents the synthesis of the compound without oxygen deficiency. One possible reason might be the slight oxydization of Sm during the synthesis in the evacuated silica tube. 
Above $x=0.075$, the lattice constants decrease monotonically with increasing $x$, similarly to $R$FeAsO$_{1-x}$. \cite{eisaki} Among all the samples synthesized in the present work, the $x$=0.075 has the largest lattice constants. Since the lattice constants for $x=0.075$ are smaller than the reported values for $x=0$, the present $x$=0.075 compound is considered to be slightly oxygen-deficient. We tried to synthesize the compounds with larger lattice constants than $x$=0.075, but could not obtain such a sample.\par

\begin{figure}[htbp]
\begin{center}
\includegraphics*[width=8cm]{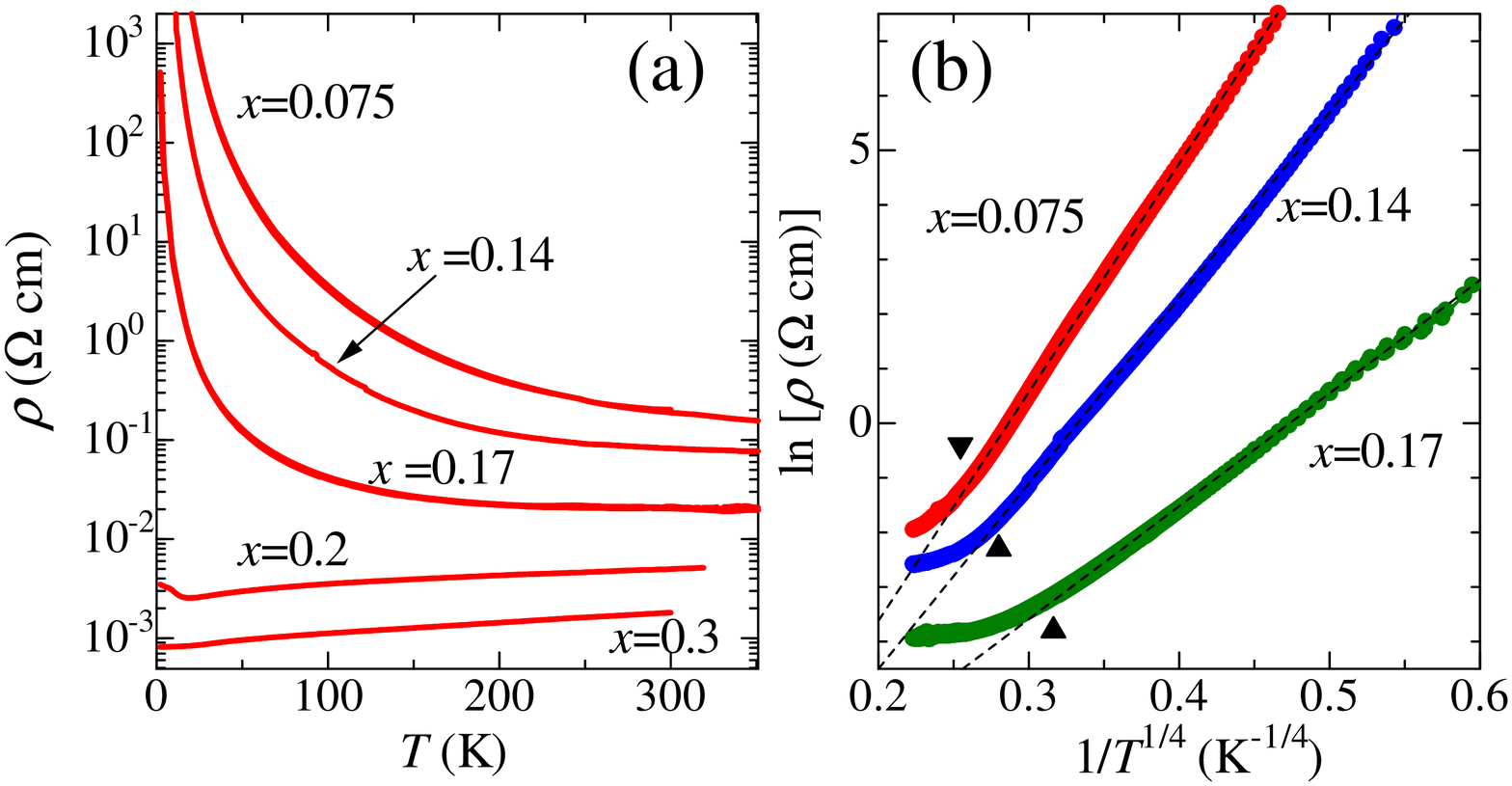}
\end{center}
\caption{(color online). (a) The temperature ($T$) dependence of resistivity ($\rho$). (b) The resistivity for the insulating samples plotted in the form of variable range hopping (VRH) relation in $3$-dimension. The dashed lines are linear fits to the data. The triangles indicate the temperature at which the VRH conduction sets in ($T_{\rm VRH}$).}
\label{fig2}
\end{figure}

\subsection{Resistivity and Seebeck coefficient}

  Figure \ref{fig2}(a) shows the temperature ($T$) dependence of the resistivity ($\rho$) for all the samples. For $x=0.075$, the temperature dependence of the resistivity is insulating and increases rapidly with decreasing temperature. It was reported for NdMnAsO that the resistivity increases with decreasing temperature above $250$ K, but tends to decrease below that temperature. \cite{Marcinkova} In contrast to such complex temperature-dependence in NdMnAsO, the resistivity of SmMnAsO$_{1-x}$ with $x=0.075$ seems to show a simple insulating behavior like LaMn$Pn$O ($Pn$= P, As, or Sb). \cite{kayanuma} As $x$ is increased, the resistivity decreases monotonically, but shows the insulating temperature-dependence up to $x=0.17$. The temperature dependence of resistivity for $x=0.075$, $0.14$, and $0.17$ samples does not follow the thermal activation type relation, 
\begin{equation}
\ln \rho(T) \propto E_{0}/k_{B}T \ \ ,
\end{equation}
but well obeys the variable range hopping (VRH) relation in three dimension, as exemplified in Fig. 2(b). The Mott's VRH relation is given by 
\begin{equation}
\ln \rho (T) \propto (T_{0}/T)^{1/4},
\label{VRH}
\end{equation}
where $T_{0}$ depends inversely on the localization length and the density of localized states at the Fermi energy. The apparent three-dimensional conduction observed in the present samples may indicate that the inter-layer hopping is substantial in this temperature range despite the layered structure. The slope of the line in Fig. \ref{fig2}(b) ($= T_{0}^{1/4}$) decreases with increasing $x$, indicating that the localization length becomes longer and/or the density of localized states increases. The temperatures where the resistivity deviates from the VRH conduction ($T_{\rm VRH}$) are indicated by black triangles in Fig. 2(b). The $x$ dependence of $T_{\rm VRH}$ is shown in Fig. \ref{fig4} (b). With increasing $x$, the VRH conduction is observed in the lower temperature region, signaling the approach to the metallic phase. \par  

As the doping proceeds to $x=0.2$, the temperature dependence of the resistivity changes to a metallic behavior. In Fig. \ref{fig4} (a), we plot the conductivity ($\sigma = 1/ \rho$) at $2$ K. The insulator-to-metal transition occurs between $x=0.17$ and $0.2$. For $x=0.2$, the resistivity decreases with decreasing temperature down to $20$ K, but it turns to increase at around that temperature. As will be discussed later, the antiferromagnetic transition of the Mn spins is observed at $27$ K, which is slightly higher than the temperature of the resistivity upturn (see also Fig. \ref{fig8}(a)). The resistivity upturn is probably due to the change in the Fermi surface shape and/or spin-dependent renormalization of transfer interaction upon the antiferromagnetic or spin density wave (SDW) transition of Mn electrons. Another possibility would be the Anderson localization in this polycrystalline sample with the disorder (oxygen deficiency), but this possibility would be ruled out from the result of magnetoresistance ({\it vide infra}). The resistivity upturn is not observed for $x=0.3$ down to $2$ K and the value is comparable with that of superconducting $R$FeAsO$_{1-x}$ \cite{eisaki}. While the residual resistivity is less than $1$ m${\rm \Omega}$cm, the superconductivity is not observed down to $0.35$ K. 
  \par

\begin{figure}[b]
\begin{center}
\includegraphics*[width=8cm]{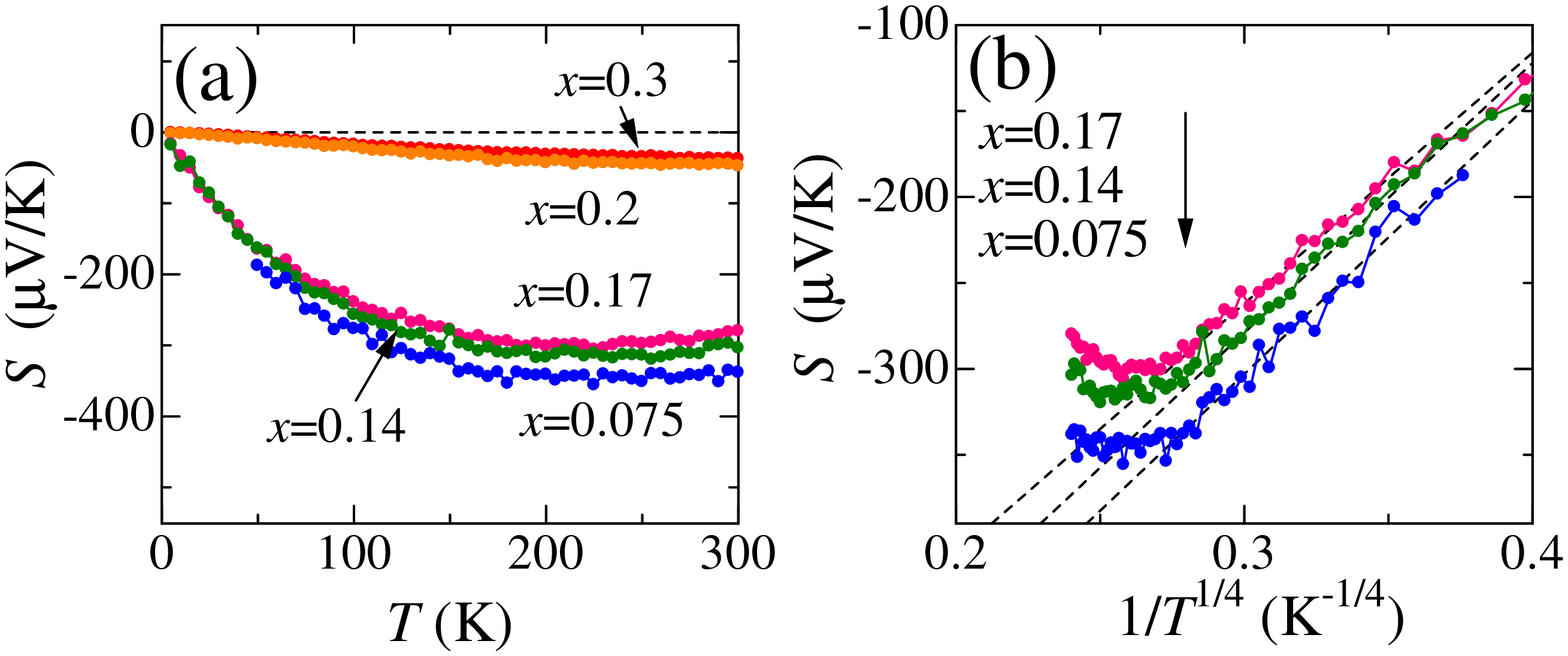}
\end{center}
\caption{(color online). (a) The temperature ($T$) dependence of the Seebeck coefficient ($S$). (b) Seebeck coefficients of insulating samples are plotted in the form of $S$ versus $1/T^{1/4}$. }
\label{fig3}
\end{figure}

Figure \ref{fig3} shows the seebeck coefficients of all the samples. As shown in Fig. \ref{fig3}(a), the Seebeck coefficient ($S$) exhibits quite a different behavior between the insulating ($x=0.075$, $0.14$, and $0.17$) and the metallic ($x=0.2$ and $0.3$) samples, while the sign of the Seebeck coefficient is negative for all the compounds. For $x=0.075$, the Seebeck coefficient increases in absolute magnitude and saturates around $-350$ ${\rm \mu}$V/K above $200$ K, and the behavior for $x=0.14$ and $0.17$ is similar to $x=0.075$ compound. As displayed in Fig. 2(b), the temperature dependence of the resistivity for these samples is well described with the VRH model. In the temperature region where charge transport is governed by VRH, the Seebeck coefficient should follow the relation, \cite{Ang} 
\begin{equation}
S(T) \propto T^{-1/4}.
\end{equation} 
As shown in Fig. \ref{fig3}(b), the Seebeck coefficient of insulating materials in the low temperature region is in good accord with this relation. The temperatures below which VRH conduction is observed are almost the same as those in the resistivity shown in Fig. 2(b). On the other hand, for the metallic samples, the absolute value of the Seebeck coefficient is much smaller than those of the insulating samples, increases monotonically with increasing temperature, and takes about $-50$ ${\rm \mu}$V/K and $-40$ ${\rm \mu}$V/K at $300$ K for $x=0.2$ and $0.3$, respectively. These values are definitely smaller than those of insulating samples, but relatively large for a metal. Because the Seebeck coefficient is less sensitive to the tiny amount of impurity phase than the resistivity, the large difference between the samples with $x \leq 0.17$ and $x \geq 0.2$ indicates that the insulator-metal transition observed in the resistivity measurement (Fig. 2(a)) is of bulk intrinsic nature. We summarize the change of the conductivity at $2$ K and the Seebeck coefficient at $300$ K across the transition in Fig. \ref{fig4} (a).    
\par

\par

\begin{figure}[htbp]
\begin{center}
\includegraphics*[width=7cm]{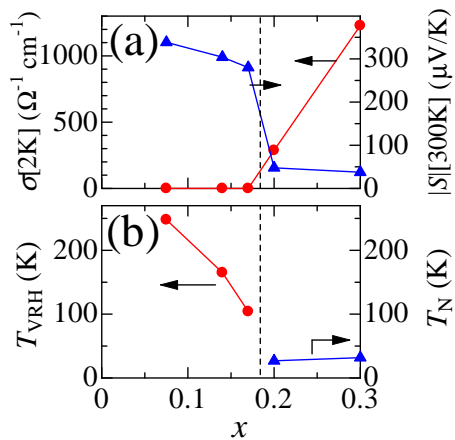}
\end{center}
\caption{(color online). Doping dependence of several physical properties. (a) The $x$ dependence of the electrical conductivity ($\sigma=1/\rho$) at $2$ K and Seebeck coefficient at $300$ K. Insulator-metal transition takes place at around $x=0.17-0.2$. (b) The $x$ dependence of $T_{\rm VRH}$ (red circles) for $x=0.075$, $0.14$, and $0.17$. $T_{N}$ of Mn spin (blue triangles) is also plotted for the metallic compounds $x=0.2$ and $0.3$. Solid lines are merely the guides to the eyes. }
\label{fig4}
\end{figure}

\subsection{Magnetization and specific heat}
                
\begin{figure}[b]
\begin{center}
\includegraphics*[width=8cm]{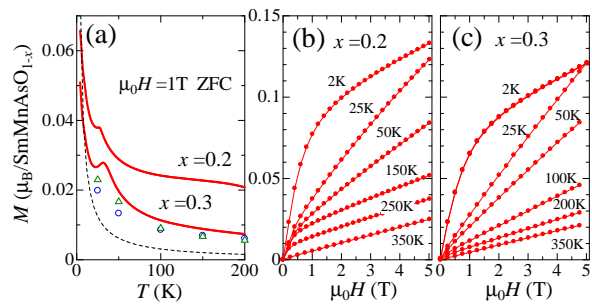}
\end{center}
\caption{(color online). (a) The temperature ($T$) dependence of the magnetization ($M$) for $x=0.2$ and $0.3$. Open circles and triangles indicate the intrinsic magnetization values of $x=0.2$ and $0.3$ estimated from the slopes of the linear parts of the isothermal $MH$ curves ((b),(c)), respectively. The detailed procedure of the estimation is described in the text. (b),(c) The magnetic-field ($H$) dependence of the magnetization ($M$) for (b) $x=0.2$ and (c) $0.3$ at several temperatures.}
\label{fig5}
\end{figure}

Figure \ref{fig5}(a) shows the magnetization ($M$) as a function of temperature for $x=0.2$ and $0.3$. For $x=0.2$, the increase of magnetization is observed around $320$ K, which is also discerned for insulating samples ($x=0.075$, $0.14$, and $x=0.17$). Since the onset temperature of this increase in magnetization does not vary with the change in $x$ and the possible impurity MnAs is known to be a ferromagnet with $T_{c} =320$ K, this magnetic transition is not an intrinsic property of SmMnAsO$_{1-x}$, but should be ascribed to a tiny amount of impurity phase (MnAs). \par

Figures \ref{fig5}(b) and (c) show the magnetic-field ($H$) dependence of magnetization for $x=0.2$ and $0.3$. For $x=0.2$, the $H$ dependence is linear above $350$ K, but below $300$ K, spontaneous moment due to the ferromagnetic impurity MnAs is observed below $1$ T. For $x=0.3$, the magnetization is almost linear in magnetic field in all the temperature region except for $2$ K. At $2$ K, the magnetization for $x=0.2$ and $0.3$ shows non-linear $H$ dependence, which should be attributed to the tiny amount of unknown Curie spins.
\par

Despite the presence of ferromagnetic component (Figs. \ref{fig5}(b) and (c)), the linear slopes in the high-$H$ region are considered to represent the intrinsic susceptibilities of $x=0.2$ and $0.3$. The intrinsic magnetizations at $1$ T are estimated from this high-$H$ linear portion above $25$ K, and plotted in Fig. \ref{fig5}(a) with the open circles and triangles, respectively. The estimated magnetizations of these compounds are almost identical above $100$ K and increase with decreasing temperature. A dahed line indicates Curie contribution ($C/T$) from Sm$^{3+}$, which is estimated from the magnetization of SmZnAsO synthesized with the same procedure using the high-pressure synthesis technique. The temperature dependence of the magnetization for $x=0.2$ and $0.3$ appears almost parallel with this Curie contribution above $50$ K, indicating the dominant contribution of Sm$^{3+}$ $f$-moment to the magnetic susceptibility. The positive offset from the magnetization of SmZnAsO is the contribution of Mn-spins which seems to be almost independent of temperature.  
\par

In the $MT$ curves shown in Fig. \ref{fig5}(a), peak structures are observed around $30$ K both for $x=0.2$ and $0.3$. Below this temperature, the magnetization slightly decreases, but increases again with decreasing temperature. This transition corresponds to the antiferromangetic or SDW order of Mn spins. The values of $T_{N}$ for $x=0.2$ and $0.3$, plotted in Fig. \ref{fig4} (b), do not differ very much from each other, and there appears to be an antiferromagnetic metallic phase prevailing next to the antiferromagnetic insulating phase. The wide range of antiferromagnetic metal phase is similar to the case of NiS$_{2-x}$Se$_{x}$. \cite{miyasaka}. On the other hand, since LaMnPO and NdMnAsO are antiferromagnets with $T_{N} > 375$ K \cite{Yanagi} and $T_{N}=359$ K, \cite{Marcinkova} respectively, SmMnAsO is likely to be an antiferromgnetic insulator with such a high $T_{N}$. No clear magnetic transition was observed for $x=0.075$ in the temperature range $300-400$ K, suggesting that the $T_{N}$ of SmMnAsO may be even higher than $400$ K. 
\par

\begin{figure}[b]
\begin{center}
\includegraphics*[width=7cm]{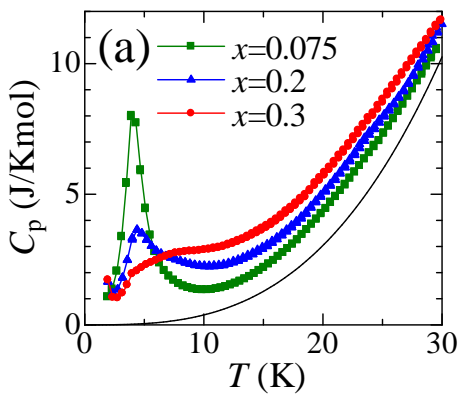}
\end{center}
\caption{(color online). (a) Temperature ($T$) dependence of specific heat ($C_{p}$) for $x=0.075$, $0.2$, and $0.3$ below $30$ K. The solid line indicates the phonon specific heat ($\beta T^{3}$). }
\label{fig6}
\end{figure}

In Fig. \ref{fig6}, we show the specific heat ($C_{p}$) in the low temperature region without magnetic field for $x=0.075$, $0.2$, and $0.3$. The peak structures are observed around $4$ K for all the samples. These anomalies are ascribed to the magnetic transitions of Sm$^{3+}$ $f$-moment. A similar peak structure was also observed around $4-5$ K in SmFeAs(O$_{1-x}$F$_{x}$), \cite{Tropeano, Riggs} and was interpreted as resulting from the antiferromagnetic transition of Sm$^{3+}$ $f$-moment on the basis of the comparison with the antiferromagneitc transition in Sm$_{2}$CuO$_{4}$ at $T_{N}=5.9$ K. \cite{Ghamaty} For the present $x=0.2$ and $0.3$, the specific heat begins to increase below $3$ K with decreasing temperature. This probably originates from the SmAs impurity, as it has been reported to show the peak structure at $2$ K in literature. \cite{LB} This assumption is consistent with the absence of the upturn below $3$ K in the $x=0.075$ compound which does not contain SmAs.
\par

The specific heat generally consists of the contributions from conduction electron ($C_{el}$), phonon ($C_{ph}$), and magnetic excitation ($C_{mg}$). $C_{el}$ and $C_{ph}$ are expressed as $C_{el} = \gamma T$ and $C_{ph} = \beta T^{3}$ at the low temperatures, where $\gamma$ and $\beta$ are electronic and phonon specific heat coefficients, respectively. As for the value of $\beta$, it is useful to refer to the result ($\beta \sim 0.36$ mJ/mol K$^{4}$) of SmFeAs(O$_{1-x}$F$_{x}$) with the same structure \cite{Tropeano, Riggs}. We estimate the $C_{mg}$ of Sm$^{3+}$ $f$-moment for $x=0.075$ by $(C_{p} - \beta T^{3})/T$ and calculate the corresponding entropy. The entropy shows a sharp increase around $4$ K and gradually approaches to $R\ln 2$, similarly to the case of SmFeAs(O$_{1-x}$F$_{x}$). As $x$ is increased, the peak structure of $C_{p}$ becomes smaller, and the magnetic contribution of Sm$^{3+}$ $f$-moment seems to be present up to the higher temperatures. Although the Mn spins order antiferromagnetically around $27$K for $x=0.2$ as described above, the corresponding anomaly is quite small. The determination of the genuine value of $\gamma$ was difficult for the present samples due to the large magnetic contribution of Sm$^{3+}$ $f$-moment.  \par

\subsection{Hall coefficient and magnetoresistance}

\begin{figure}[b]
\begin{center}
\includegraphics*[width=8cm]{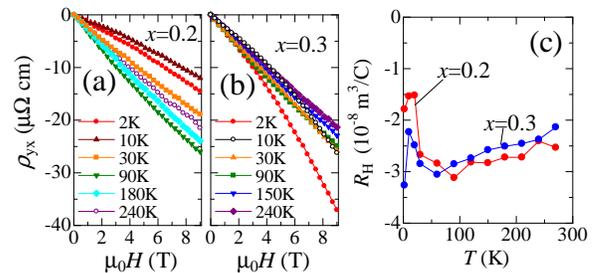}
\end{center}
\caption{(color online). (a),(b) The magnetic field ($H$) dependence  of Hall resistivity for (a) $x=0.2$ and (b) $0.3$. The contribution arising from the magnetoresistance of the longitudinal resistivity is eliminated by using the Hall voltage values at $\pm H$. (c) The temperature ($T$) dependence of the Hall coefficient ($\equiv R_{H}$). }
\label{fig7}
\end{figure}

In order to obtain the information on the Fermi surface in the metallic samples ($x=0.2$ and $0.3$), we measured the Hall resistivity $\rho_{yx}$ (Figs. \ref{fig7}(a) and (b)). The slope of the Hall resistivity is negative both for $x=0.2$ and $0.3$, and almost linear with magnetic field above $T_{N}$. In this temperature region, the Hall coefficient gradually increases in absolute magnitude as the temperature is decreased, as plotted in Fig. \ref{fig7}(c). 
The observed absolute value of the Hall coefficient is one order of magnitude larger than that expected from the nominal oxygen deficiency. This fact can be accounted for in terms of the existence of the hole-type Fermi surface in addition to the electron-type one in the metallic phase derived from the Mott insulator, where a dramatic reconstruction of the band structure is expected. According to the band calculation for $R$FeAsO, \cite{kuroki,Mazin,Pickett,Terakura} the multiband structure consisting of five  Fe-$3d$ bands is suggested near the Fermi energy and there seem to be both electron and hole pockets as the Fermi surfaces. Although the band calculation has not been performed for the metallic SmMnAsO$_{1-x}$, the complex multiband structure is inferred as well from the present result of Hall coefficient.


 \par

The magnitude of the slope of the $\rho_{yx}$ v.s. $H$ curve in the low-field region begins to decrease at $30$ K both for $x=0.2$ and $x=0.3$. Since the antiferromagnetic transition occurs around $30$ K, the decrease of the absolute value of Hall coefficient ($R_{H}$) should be related with the change in the electronic structure. However, the absolute magnitude of $R_{H}$ usually increases below the antiferromagnetic transition because the Fermi surface shrinks and the carrier density decreases. \cite{uchida, miyasaka} Again, the complex multiband structure of Mn-$3d$-bands around the Fermi energy would be the origin for the apparent decrease in the absolute magnitude of the Hall coefficient. The $R_{H}$ in the presence of both electron- and hole-type carriers is generally expressed as \cite{smith}
\begin{equation}
R_{H} = \frac{1}{e} \frac{(n_{h}\mu_{h}^{2} - n_{e}\mu_{e}^{2} ) + \mu_{e}^{2}\mu_{h}^{2}(n_{h}-n_{e}) (\mu_{0}H)^{2} }{ (\mu_{e}n_{e} + \mu_{h}n_{h} )^{2} + \mu_{e}^{2}\mu_{h}^{2}(n_{h}-n_{e})^{2} (\mu_{0}H)^{2}  } , 
\label{hall}
\end{equation} 
where $n_{e}$ ($n_{h}$) and $\mu_{e}$ ($\mu_{h}$) are density and mobility of electron (hole), respectively. The $H^{2}$-term in Eq.(\ref{hall}) is significant only in the low temperature regions of good metals with high mobility, which should be far from the present case with substantial disorders and hence with poor mobility. Therefore, we can safely use the formula in the low-field limit, $R_{H}= e^{-1} \cdot (n_{h} \mu_{h}^{2} - n_{e} \mu_{e}^{2})/ (n_{e}\mu_{e} + n_{h}\mu_{h})^{2}$. Upon the SDW transition, a gap is likely to open at the electron-type Fermi surface, which decreases $n_{e}$ and hence the magnitude of negative $R_{H}$ for $x=0.2$. The Hall resistivity $\rho_{yx}$ shows convex upward curvature below $T_{N}$. This behavior is ascribed to the field-induced change of electronic structure or related parameters ($n_{e}$, $\mu_{e}$, $n_{h}$, $\mu_{h}$), which also manifests itself as the unconventional large positive magnetoresistance as described in the following. 

\par

\begin{figure}[htbp]
\begin{center}
\includegraphics*[width=7cm]{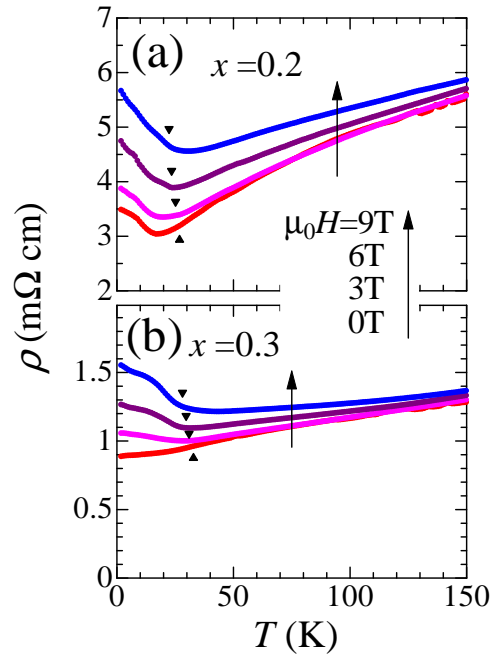}
\end{center}
\caption{(color online). (a),(b) The temperature ($T$) dependence of resistivity ($\rho$) for (a) $x=0.2$ and (b) $x=0.3$ in various magnetic fields. The triangles indicate the $T_{N}$ determined from the magnetization data.  }
\label{fig8}
\end{figure}

 Figures \ref{fig8}(a) and (b) show the temperature dependence of the resistivity in several magnetic fields up to $9$ T for $x=0.2$ and $0.3$, respectively. In both compounds, the unexpectedly large positive magnetoresistance effect is observed in a wide range of temperature and its magnitude increases with decreasing temperature. The triangles in Figs. \ref{fig8}(a) and (b) indicate the values of $T_{N}$ in magnetic fields, which are determined from the magnetization data. $T_{N}$ value slightly decreases both for $x=0.2$ and $0.3$ as the magnetic field is increased. For $x=0.2$, the resistivity shows an upturn below a temperature slightly above $T_{N}$ in zero field, and the temperature of the resistivity minimum increases with increase in the magnetic field. For $x=0.3$, the resistivity monotonically decreases down to the lowest temperature in zero field, and the temperature-dependence is proportional to $T^{2}$ below $25$ K. In magnetic fields above $3$ T, the resistivity upturn becomes visible below $T_{N}$. \par

\begin{figure}[t]
\begin{center}
\includegraphics*[width=8cm]{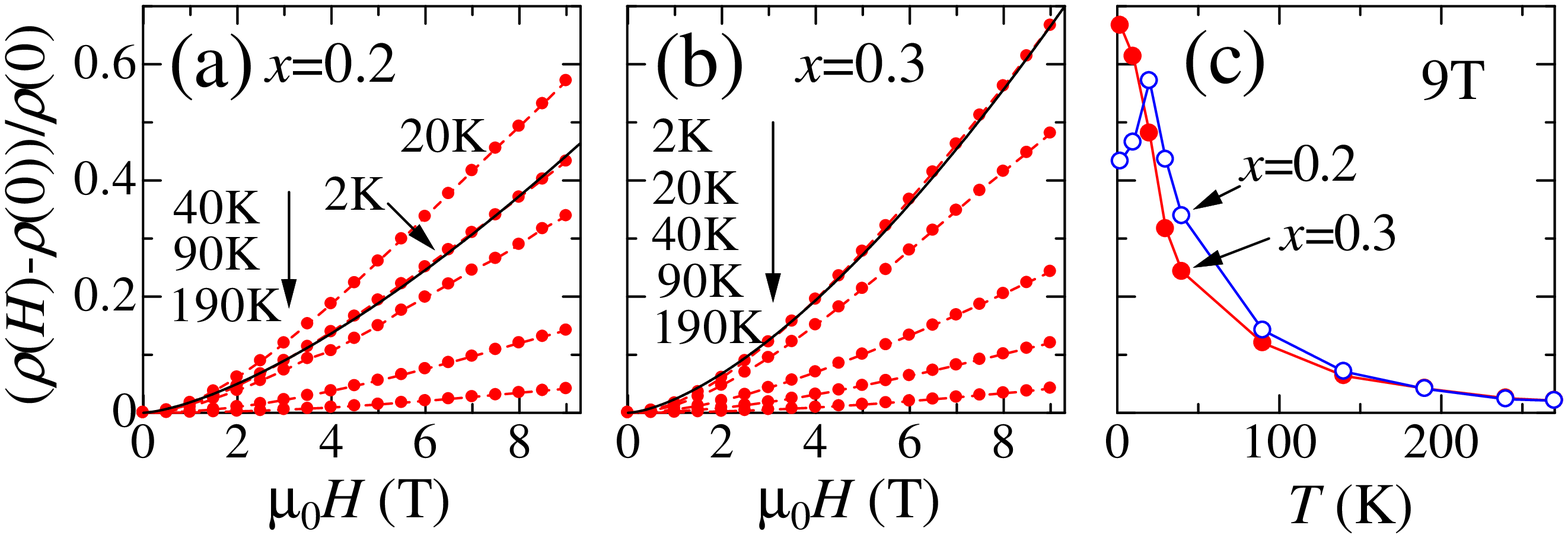}
\end{center}
\caption{(color online). Isothermal magnetoresistance magnitude [$ \equiv \{ \rho(H)-\rho(0) \} /\rho(0)$] for (a) $x=0.2$ and (b) $x=0.3$ at various temperatures. The solid lines indicate the power-low dependence of $H^{1.5}$. (c) The temperature ($T$) dependence of the (longitudinal) magnetoresistance magnitude at $9$ T for $x=0.2$ (open circles) and $0.3$ (closed circles).  }
\label{fig9}
\end{figure}

In Figs. \ref{fig9}(a) and (b), we show the field-dependence of the magnetoresistance magnitude [$\{ \rho(H)-\rho(H=0) \} /\rho(H=0)$]. The magnetoresistance magnitude is almost proportional to $H^{1.5}$ up to $9$ T both for $x=0.2$ and $x=0.3$ at all the temperatures, as exemplified by the solid curves in Figs. \ref{fig9}(a) and (b). We have tried the Kohler's plot (not shown), but these magnetoresistance data at several temperatures from $2$ K to $200$ K do not collapse on a single curve either for $x=0.2$ or $x=0.3$. Usually, violation of the Kohler's rule indicates that the scattering rates for electron- and hole-type carriers are different, and/or that the scattering rates are non-uniform over the Fermi surface. In the present case, however, this should be interpreted as an indication of different mechanism(s) of magnetoresistance from the bending of electron orbital. In Fig. \ref{fig9}(c), magnetoresistance magnitude at $9$ T is plotted against temperature. The magnetoresistance shows a maximum at $20$ K (slightly below $T_{N}$) for $x=0.2$, while it increases monotonically down to the lowest temperature for $x=0.3$. The magnitude of the magnetoresistance seems to be little dependent on the value of $x$ above $T_{N}$, reaches more than $50$ \% at the maximum, and persists to high temperatures, {\it e.g.} $5$ \% even at $190$ K.
\par

As possible origins of such large magnetoresistance, we may consider the following five mechanisms; (i)grain-boundary scattering, (ii)weak localization, (iii)spin scattering, (iv)Lorentz force, and (v)changes in the carrier density and/or the mobility associated with the change in the band structure. The mechanism (v) would be most plausible as the origin for the large positive magnetoresistance, although we cannot uniquely identify the origin. In the following, we briefly discuss these mechanisms one by one; (i) The large {\it negative} magnetoresistance associated with the spin-polarized intergrain tunneling is observed at low temperatures in a poly-crystalline ferromagnetic metal of La$_{2/3}$Sr$_{1/3}$MnO$_{3}$. \cite{hwang} Apart from the sign, the temperature dependence shown in Fig. \ref{fig9}(c) is reminiscent of this intergrain-tunneling magnetoresistance. However, SmMnAsO$_{1-x}$ is antiferromagnetic and the tunneling process should not be affected by the relative angle of the spin directions of the neighboring grains. Indeed, similar positive magnetoresistance has been observed both for single crystal and polycrystalline sample of antiferromagnetic $R$FeAsO. \cite{jdong, mcguire, cheng}
\par

(ii) The upturn of the resistivity in the zero field for $x=0.2$ may be related with the weak-localization effect. In that case, the field effect on the localization would manifest the magnetoresistance. However, as shown in Figs. \ref{fig9}(a) and (b), the $H^{1.5}$-dependence of the magnetoresistance magnitude is clearly different from $H^{0.5}$ or $\ln H$ behavior expected for weak localization, and does not change across the temperature of the resistivity upturn. Therefore, the weak localization should be irrelevant to the observed magnetoresistance effect.
\par
 
(iii) In view of the fact that the magnetoresistance magnitude takes a maximum around $T_{N}$ for $x=0.2$, the magnetoresistance would be somehow related with the antiferromagnetic transition of Mn spins. The field-suppression of the spin scattering is known to be a major source of the magnetoresistance in magnetic metals, but it usually causes a {\it negative} magnetoresistance effect, \cite{usami} as observed in the SDW states of metallic V$_{2-y}$O$_{3}$. \cite{klimm} 
\par

(iv) Positive magnetoresistance is generally caused by the Lorentz force. \cite{MR-note} The magnetoresistance in the presence of both electron- and hole-types carriers is expressed as \cite{ziman}\\
\hspace{-1cm}
\begin{minipage}[c]{10cm}
\begin{equation}
\frac{\rho(H)- \rho(0)}{\rho(0)} = \frac{n_{e}n_{h} \mu_{e} \mu_{h} (\mu_{e} - \mu_{h})^{2}  (\mu_{0}H)^{2}}{(n_{e}\mu_{e} + n_{h} \mu_{h})^{2} + (\mu_{0}H)^{2} (\mu_{e}\mu_{h}n_{h} + \mu_{h}\mu_{e}n_{e})^{2} }.
\label{MR}
\end{equation}
\end{minipage}\\
The $H^{1.5}$-dependence of the magnetoresistance magnitude might be viewed as close to the $H^{2}$-dependence represented by Eq.(\ref{MR}). However, the magnitude of the magnetoresistance induced by the Lorentz force is usually as small as a few percent, except for metals with very high mobility. The present materials are strongly correlated metals derived from a Mott insulator, and also contain substantial disorder. Therefore, it would be unlikely that Lorentz force gives rise to such a gigantic magnetoresistance as exceeding $50$ \%.
 \par

(v) Similar positive magnetoresistance effect as large as several tens of percent has been reported in SDW states of poly- and single-crystalline $R$FeAsO \cite{jdong, mcguire, cheng} and single-crystalline SrFe$_{2}$As$_{2}$ \cite{chen}. For the present SmMnAsO$_{1-x}$, as inferred from the Hall effect data, gap opening on the electron Fermi surface takes place upon the SDW transition. Also, the slope of the Hall resistivity becomes large as the field is increased, as shown in Figs. \ref{fig7}(a) and (b). Considering the relation $R_{H}= e^{-1} \cdot (n_{h} \mu_{h}^{2} - n_{e} \mu_{e}^{2})/ (n_{e}\mu_{e} + n_{h}\mu_{h})^{2}$, this observation indicates the increase in $n_{e}$ and/or $\mu_{e}$, or the decrease in $n_{h}$ and/or $\mu_{h}$. Although the mechanism is not clear at present, if the $n_{h}$ and/or $\mu_{h}$ is reduced by the application of the magnetic field, then the positive magnetoresistance as well as the field dependence of the Hall resistivity can be explained in a consistent manner. This situation reminds us of the case of ferromagnetic (Fe,Co)Si \cite{onose}, where the large positive magnetoresistance effect was observed in the low temperature region. In this case, the Zeeman shift of exchange split bands decrease the carrier density in the minority band, leading to the increase of the total resistivity, provided that the mobility of the minority-spin band is larger than that of the majority-spin band. In the case of SmMnAsO$_{1-x}$, electron- and hole-type carriers might play the similar role to the case of majority and minority spin-band, respectively, and the change in the Fermi surface as affected also by the Zeeman splitting would be important. However, the situation in the present case is more complex than in the (Fe,Co)Si case, due to the presence of four kinds of carriers, {\it i.e.} electron- and hole-type carriers with up and down spin; this makes the quantitative analysis difficult. 

 \par

\section{Conclusions}
We synthesized a series of SmMnAsO$_{1-x}$ with controlled electron-doping level by using high-pressure synthesis technique. We showed that, as the electron-doping is increased with increasing oxygen deficiency, the resistivity decreases monotonically and the insulator-metal transition occurs between $x$=0.17 and $0.2$. The temperature dependence of the Seebeck coefficient is largely different between $x \leq 0.17$ and $x \geq 0.2$, confirming the insulator-metal transition of bulk nature in accord with the resistivity measurement. The metallic $x=0.2$ and $0.3$ compounds show the antiferromagnetic order at low temperatures below $30$ K. Upon the antiferromagnetic or SDW transition, the metallic compounds show a reduction in the magnitude of Hall coefficient (with negative sign) in low-field region, indicating the coexistence of electron- and hole-type carriers as well as the gap opening on the electron-type Fermi surface. Both of the $x=0.2$ and $0.3$ compounds show the unusually large positive magnetoresistance effect in a wide range of temperature, which exceeds $50$ \% at $2$ K. We attribute the observed magnetoresistance effect also to the field-induced change in the band structure consisting of complex multiple Fermi-surfaces.

\begin{acknowledgments}
The authors would like to thank W. Koshibae and J. Checkelsky for useful discussions, and K. Ohgushi and T. Suzuki for their technical helps. One of the authors (Y.S.) was supported by the Junior Research Associate (JRA) program at RIKEN. This work was in part supported by Grant-In-Aid for Scientific Research (Grant No. 20340086.) and JSPS Fellows, and by the Funding Program for World-Leading Innovative R\&D on Science and Technology (FIRST program) on ``Quantum Science on Strong Correlation".
\end{acknowledgments}



\begin{thebibliography}{99}
\bibitem{MR-note}In our measurement geometry with the magnetic field parallel to the current, the resistivity should not be affected by the Lorentz force basically. However, if the electron flow is almost limited along the in-plane direction of randomly-oriented each grain in poly-crystalline SmMnAsO$_{1-x}$, the Lorentz force would cause the magnetoresistance.
\end{thebibliography}
\end{document}